\documentstyle[aps,prl,twocolumn]{revtex}
\input{psfig}

\def \vs {{\bf v_s}}
\def \k {{\bf k}}
\def \r  {{\bf r}}
\def \d  {{\rm d}}

\begin{document}

\bibliographystyle{unsrt}
\newcommand{\be}{\begin{equation}}
\newcommand{\ee}{\end{equation}}
\newcommand{\bea}{\begin{eqnarray}}
\newcommand{\eea}{\end{eqnarray}}

\twocolumn[\hsize\textwidth\columnwidth\hsize\csname@twocolumnfalse%
\endcsname

\title
{Magnetic field as a probe of gap energy scales in YBCO}

\author{N. D. Whelan and J. P. Carbotte}
\address{Department of Physics and Astronomy, McMaster University,
Hamilton, Ontario, Canada L8S~4M1}

\date{\today}

\maketitle

\begin{abstract}
Among high T$_c$ materials, the YBCO (YBa$_2$Cu$_3$O$_{7-x}$)
compounds are special since they have superconducting chains as well
as planes. We show that a discontinuity in the density of states as a
function of magnetic field may appear at a new energy scale,
characteristic of the chain and distinct from that set by the $d$ wave
gap. This is observable in experimental studies of the thermodynamical 
properties of these systems, such as the specific heat.
\end{abstract}


]

It is widely accepted that a $d$-wave pairing gap describes the
superconductivity in high T$_c$ compounds\cite{dwave}.  The YBCO
compounds are special however, due to the fact that the one
dimensional chains change the Fermi surface, giving a quasi-linear
branch, and thereby introducing a new energy scale for the variation
of the gap on the Fermi surface. We argue that this can be directly
measured using a magnetic field.

Scanning tunnelling microscopy (STM) measurements \cite{stm1,stm2} of
the local density of states of YBCO agree with earlier tunnelling
measurements \cite{dv} and show a very clear structure around $20$ meV
corresponding to the full gap energy. They also indicate structure at
around $5.5$ meV which is absent in BSCCO compounds \cite{stm2}. It is
natural to assume that this is because YBCO has a different electronic
structure, possessing orthorhombic chains as well as tetragonal
planes. The chains introduce a new energy scale, the value of the gap
function at the endpoint of the chain Fermi surface on the boundary of
the Brillouin zone. Since experiments are only sensitive to the value
of the gap function {\it on} the Fermi surface (and not to its global
behaviour throughout the Brillouin zone) this energy scale will be
apparent in experiments. Another interpretation of the additional
structure is localised quasi-particle states in the vortex core
\cite{caroli}. Even if present, these should have a negligible effect
on the bulk density of states, a point also made in Ref.~\cite{kh}.

We model this by assigning the chains an elliptical Fermi surface but
with a highly modified gap function
\be \label{sdgap}
\Delta_\k = \Delta_0 (\cos 2\phi + s).
\ee
$\phi$ is the polar coordinate in $\k$ space and the dimensionless
parameter $s$ (with $|s|<1$) indicates the effective amount of $s$
wave in the order parameter. We stress that (\ref{sdgap}) is not the
gap function defined throughout the Brillouin zone but is rather its
value evaluated on the Fermi surface. Along the chain Fermi surface
the gap function will have a dependence like (\ref{sdgap}) even if it
is globally pure $d$-wave. This is an effective theory to model the
contribution of the chains to the density of states and reflects a
combination of gap symmetry and Fermi surface geometry. (Independent
information on the existence of a subdominant $s$-wave component is
provided by Josephson tunnelling data
\cite{dswave,ddom}.)

\begin{figure}[h]
\hspace*{0.0in}\psfig{figure=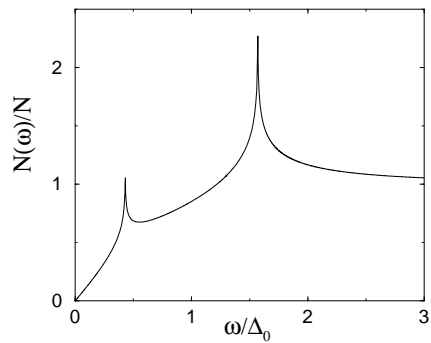,height=1.8in}
\caption
{\small Typical result for $N(\omega)$ for a pure superconductor using
$s=0.57$ to introduce a second low energy scale. Note the van Hove
singularities at $\omega/\Delta_0=(1\pm |s|)$.}
\label{nofomega}
\end{figure}

Standard theory for the density of states gives \cite{standard}
\be \label{standard}
{N(\omega)\over\bar{N}} \propto \int \d\k \; {\rm Im} \left\{
{|\omega|\over \omega^2-\Delta_\k^2-\xi_\k^2}\right\},
\ee
where $\xi_\k$ is the dispersion relation for the quasi-particle
excitation energy (for the moment we take it $\xi_k=\k^2/2m$ with $\k$
the momentum and $m$ the effective mass) and $\bar{N}$ is the normal,
nonsuperconducting density of states. Using the gap function
(\ref{sdgap}), this integral is easily done and is plotted in
Fig.~\ref{nofomega} using $s=0.57$ (this choice is explained below.)
There is a van Hove singularity at $\omega=(1+|s|)\Delta_0$ as we
expect \cite{vanhove}, however there is another at $(1-|s|)\Delta_0$,
in qualitative agreement with \cite{stm1,stm2}.

A problem with the measured $N(\omega)$ is that there is a large zero
bias anomaly in contradiction with theory and presumably arising from
surface effects. It is therefore useful and possibly better to
measure a bulk property. We propose the low temperature limit
of the specific heat, which comes from the density of quasiparticle
states at zero frequency.
%In principle one could use the specific heat
%as a function of temperature but in practise temperature measurements
%tend to smear out structure. In any case, at higher temperatures, the
%specific heat is dominated by phonons rather than electrons.
We focus on the magnetic field regime $H_{c1}\ll H \ll H_{c2}$ where
the density of states is given by quasi-particle excitations in the
presence of the induced vortices. For each magnetic field, there is a
magnetic energy, which scales as $\sqrt{H}$ \cite{volovik,vekhter}. We
expect structure at critical fields for which the magnetic energy
sweeps through the singularities at $\Delta_0 (1\pm s)$ in
Fig.~\ref{nofomega}.  The new, lower energy scale is well within the
range of available fields and is observable in specific heat
experiments which probe the vortex modified density of states.

In the presence of the vortex, the Cooper pairs have a superfluid
velocity $\vs$ which depends inversely on the distance to the vortex
core. The quasi-particle energies of momentum $\k$ are Doppler shifted
by $\omega\rightarrow\omega + \vs\cdot\k$ \cite{tinkham}, which can be
thought of as introducing a spatially dependent shift in chemical
potential.  We are interested in the zero frequency case, so we can
use Eq.~(\ref{standard}) but with $\omega$ replaced by $V=\vs\cdot\k$.
In addition to the trace integral over $\k$, we also do a
spatial average within the vortex unit cell, with coordinates $\r$:
\be \label{initint}
N_0(H) \propto \int \d\k \int \d\r \;
               V \delta \left(\zeta_{\k}^2 - V^2 + \Delta_{\k}^2\right).
\ee

To find $\vs$, we use the free energy density per unit length
\be
F = \int \d^2r \left( B^2 + \lambda_x^2 (\nabla \times B)_x^2
                          + \lambda_y^2 (\nabla \times B)_y^2
               \right).
\ee
We have introduced the parameters $\lambda_i^2 = \mu_i \lambda^2$ with
$\lambda^2=Mc^2/4\pi e^2 n_s$, $\mu_i = m_i/M$ and
$M=\sqrt{m_xm_y}$ ($\lambda$ is a mean London penetration depth and
$n_s$ the superfluid density.) Following standard free energy
minimisation\cite{tinkham} with a flux line source at the origin, we
express the magnetic field as a modified Bessel function. In the
magnetic field regime considered here, the typical spacing between
vortices is much smaller than $\lambda$ and we use the small
argument approximation:
\be
B \approx  -{\Phi_0\over 2\pi\lambda^2} 
\log\left({1\over\lambda}\sqrt{{x^2\over\mu_y} + {y^2\over\mu_x}}\right),
\ee
pointing in the z direction and $\Phi_0=\pi\hbar c/e$ is the flux
quantum. Applying Amp\`ere's law, the current is
\bea
j_x & \approx &      -{c\Phi_0\over 8\pi^2\lambda^2} 
{1\over {x^2\over\mu_y} + {y^2\over\mu_x}} {y\over\mu_x} \nonumber\\
j_y & \approx & \phantom{-}{c\Phi_0\over 8\pi^2\lambda^2} 
{1\over {x^2\over\mu_y} + {y^2\over\mu_x}} {x\over\mu_y}.
\eea
We find $\vs$ by dividing the current by $en_s$.

The quasi-particle excitations are given by the dispersion relation
\be
\zeta_{\k} = {\hbar^2k_x^2\over 2m_x} + {\hbar^2k_y^2\over 2m_y}.
\ee
As discussed above, we take the gap function (\ref{sdgap}) to be
\be
\Delta_\k = \Delta_0 f(\k) = \Delta_0\left( {k_x^2 - k_y^2\over
k_x^2+k_y^2} + s\right),
\ee
The analysis is simplified by a change of variables to
\bea
x'  = x/\sqrt{\mu_y} \;\;\; & \;\;\; k_x'= \sqrt{\mu_y}k_x \nonumber\\
y'  = y/\sqrt{\mu_x} \;\;\; & \;\;\; k_y'= \sqrt{\mu_x}k_y.
\eea
In terms of these new coordinates we have
\bea \label{stuff}
\vs       & = & {\hbar\over 2M} {1\over r'} \hat{\beta}'\nonumber\\
\zeta_\k  & = & {\k'^2 \over 2M}\nonumber\\
f(\k)     & = & f(\phi') = 
{m_y/m_x - \tan^2\phi' \over m_y/m_x + \tan^2\phi'} + s
\eea
where $(r',\beta')$ are the spatial polar coordinates ($\beta'$ is the
vortex winding angle) and $(k',\phi')$ are the momentum polar
coordinates in the new coordinates. Note that $\vs\cdot\k =
|\vs||\k|\sin(\beta'-\phi')$. Henceforth we drop the primes. Since we
are integrating over $\beta$, we are also free to shift its origin and
thereby replace $\beta-\phi$ by $\beta$. To conform to usual notation
we say that the chains run along the $b$ direction which we define as
parallel to the $y$ axis. Since the carrier mass in the $b$ direction
is less than in the $a$ direction (due to the conductivity supplied by
the chains) we conclude that $m_y/m_x<1$. In principle $s$ can have
either sign (we ignore the time-reversal symmetry breaking possibility
for which $s$ is complex.) For purposes of exposition we will use the
realistic value $m_y/m_x=0.5$ and take $s=0.57$ so as to get a second
gap scale around $~5.5$ meV as seen in experiment. The oscillatory
part of the gap function in (\ref{stuff}) has been modified due to the
coordinate rescaling, compared to the original $\Delta_0\cos 2\phi$.
Nevertheless, the amplitude has not been modified and without the $s$
term both functions would vary between plus and minus $\Delta_0$.
Either form would yield an initial linear dependence on $\sqrt{H}$
\cite{volovik,vekhter} followed by a saturation at a scale set by
$\Delta_0$. Rather, it is the $s$ term which changes this qualitative
picture. It makes the gap function have a maximum value of
$\Delta_0(1+|s|)$ and a minimum value of $\Delta_0(1-|s|)$. The first
of these gives the saturation scale mentioned above while the second
gives structure corresponding to the first peak in
Fig.~\ref{nofomega}.

We begin by nondimensionalising the spatial integral in
(\ref{initint}). In order for the average magnetic field in the sample
to equal the applied field, we require the inter-vortex spacing to be
\cite{vekhter} $R = \sqrt{\Phi_0/\pi H}/a$ where $a$ is a geometrical
constant of order unity which accounts for the mismatch between the
circular vortices and the hexagonal lattice they fill out.  We define
the magnetic energy as
\be
E_H = {a\over 2M} v_F \sqrt{{\pi H \over\Phi_0}}
\ee
where $v_F$ is the Fermi velocity and the dimensionless magnetic
energy as $\nu = E_H/\Delta_0$. We then express (\ref{initint}) as
\be \label{middleint}
{N_0(H)\over \bar{N}} = {1\over 2\pi^2} 
\int_0^{2\pi} \d\phi \int_0^{2\pi} \d\beta \int_0^1 \d\rho
{\rho \over \sqrt{1-\left({\rho f(\phi) \over \nu \sin \beta } \right)^2}}
\ee
where $\bar{N}$ is the normal density of states and $\rho=r/R$.  We
used the delta function to perform the $k$ integral and it is
understood that the integration domain is limited to the range where
the integrand is real.

We can do the $\rho$ and $\beta$ integrals in (\ref{middleint})
leading to the final expression for the density of states,
\be \label{finalint}
{N_0(H)\over \bar{N}} = {1\over 2\pi} \int \d\phi \;{\rm min} 
\left\{1,\left({\nu\over f(\phi)}\right)^2\right\}.
\ee
The parameter $\nu$ is experimentally tunable while $s$ (which enters
$f(\phi)$ through (\ref{stuff})) is a fixed, intrinsic material
parameter (although it may depend somewhat on doping levels) and
represents a combination of gap and Fermi surface symmetries. For the
case $|s|<1$ and $\nu\ll 1$, it is simpler to go back to
(\ref{middleint}) and expand the integrand around the gap nodes. One
can perform the $\phi$ integral approximately in the limit of small
$\nu$ and the resulting $\beta$ and $\rho$ integrals are trivial so
that
\be \label{smallnu}
{N_0(H)\over \bar{N}} \approx \left({2\over\pi}\sum_n{1\over
|f'_n|}\right) \nu.
\ee
The index $n$ refers to the nodes of the gap function and $f'_n$ is
the derivative of the function $f$ evaluated at the nodes. This is
still linear, as one has for a pure $d$-wave gap, although the
presence of the $s$ term does change the positions of the gap nodes
and hence affects the slope. For $|s|>1$, there are no nodes and the
dependence is quadratic for small $\nu$. Since this is unphysical we
do not present the explicit form.

\begin{figure}[h]
\hspace*{0.0in}\psfig{figure=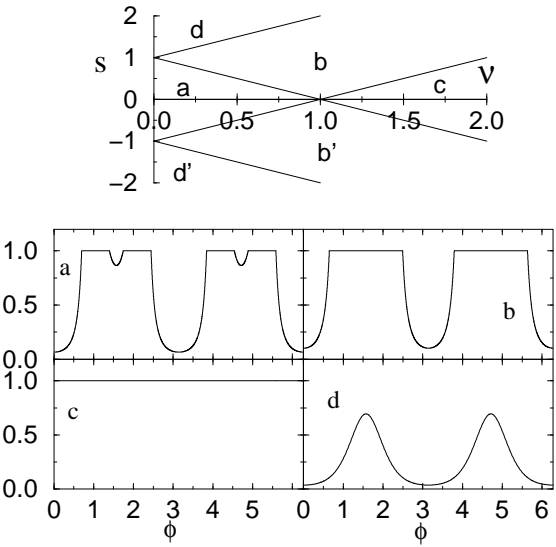,height=2.8in}
\caption
{\small The four possible behaviours of the integrand of
Eq.~(\ref{finalint}) as a function of $\phi$. The values of $(\nu,s)$
are $(0.4,0.57)$, $(0.5,0.57)$, $(2.0,0.57)$ and $(0.5,1.6)$ for cases
a to d respectively. In all cases the mass anisotropy was
$m_y/m_x=0.5$. Top: regions in the $\nu-s$ plane corresponding to the
4 types of integrands (in regions b' and d' the integrand is
translated by $\pi/2$ from its form in regions b and d.) }
\label{4graphs}
\end{figure}

\begin{figure}[h]
\hspace*{0.0in}\psfig{figure=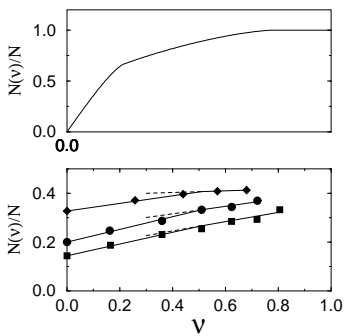,height=1.8in}
\caption{\small Top: the density of states plotted against the
rescaled magnetic field $\nu$ for the case $s=0.57$ and $m_y/m_x=0.5$.
Bottom: experimental data converted to dimensionless units as
described in the text. Circles, diamonds and squares are from
[15,16,18] respectively. In each case we have estimated a best-fit
line below and above $\nu=0.5$ to guide the eye.}
\label{nofnu}
\end{figure}

There are four distinct behaviours of the integrand of
(\ref{finalint}), as shown in Fig.~\ref{4graphs} and the $\nu-s$ plane
is accordingly divided into six regions. For fixed $s$, we vary $\nu$
and thereby cross from one region to another. Associated with such
crossings there is a corresponding nonanalyticity in the density of
states, which is experimentally measurable. This is shown in
Fig.~\ref{nofnu} where we plot the density of states as a function of
$\nu$ using the integral form (\ref{finalint}).  Clearly there is
structure at $\nu=0.43$ and $1.57$, in accord with our general
considerations. The kink at $\nu=1.57$ corresponds to the integrand
saturating at unity --- as we go from an integrand as in
Fig.~\ref{4graphs}b to the type as in Fig.~\ref{4graphs}c. However,
this probably lies beyond any realisable magnetic field strength. The
kink at $\nu=0.43$ which corresponds to going from an integrand as in
Fig.~\ref{4graphs}a to the type in Fig.~\ref{4graphs}b appears by eye
to be a discontinuity in slope. This is not really the case, as we now
discuss.

The integrand in Fig.~\ref{4graphs}a saturates around $\phi=\pi/2$ and
$\phi=3\pi/2$ when $\nu$ approaches $1-|s|$ and the integral does not
increase as quickly after this value as it does before. We can
evaluate the missing contribution, $\delta N$, from the regions around
these local minima. We take $\nu$ to be just below this transition,
{\it i.e.}
\be
\nu = 1 - |s| - \epsilon
\ee
with $\epsilon\ll 1$. To leading order in $\epsilon$, the contribution 
of this region to (\ref{finalint}) is
\be \label{discont}
{\delta N\over\bar{N}}
\approx -{8\over 3\pi (1-|s|)} \sqrt{m_x\over 2m_y}\epsilon^{3/2}.
\ee
There is no such contribution if $\nu$ is just above $1-|s|$, so there
is a discontinuity in the density of states. While the precise
prefactor and position of the discontinuity may depend on factors
which we have not included, the three-halves power is generic,
depending only on the topological property of having local minima of
the integrand disappear. For example, in the event that $s<0$, we
first lose the minimum around $\phi=0$ instead of $\phi=\pi/2$ but
otherwise the behaviour is the same and, except for a change in
prefactor (typically it is smaller), we still expect a discontinuity
of this power. Similarly, it is the same order of discontinuity at
$\nu=1+|s|$ although, as stated, this is probably beyond the
accessible magnetic field range.

To compare with experiment, we now determine the parameters of our
model. The upper scale of $20$ meV equals $(1+s)\Delta_0$ and the
lower scale of 5.5 meV equals $(1-s)\Delta_0$ from which we infer
$\Delta_0\approx 12.7 {\rm meV}$ and $s\approx 0.57$. We also use the
result, consistent with both $\mu$SR and specific heat measurements
that $E_H\approx 2.6\sqrt{H}$ when $H$ is measured in Tesla and $E_H$
in meV\cite{units}.

The dimensionless parameter $\nu$ then equals $0.20\sqrt{H}$. The
dependence of the specific heat on $H$ has been measured in
\cite{moler,fisher,junod1,sisson}, with \cite{sisson} having the
cleanest samples. The data is shown in Fig.~\ref{nofnu} in
dimensionless units (taking the specific heat at low temperatures to
be proportional to temperature and to the density of states.) By our
previous arguments, energies of $5.5$ meV correspond to magnetic
fields of about $4.5T$ or $\nu$ of about $0.42$, which is well within
the experimental ranges considered.  The origin of the finite value of
the density of states at zero field is not well understood but is
sometimes attributed to oxygen vacancies on the chain \cite{sisson} or
to resonant impurity scattering \cite{kh}.  Either way, to compare
with our results one should subtract off this constant. There will
also be a linearly increasing component from the density of states on
the planes; this should be of comparable magnitude but without the
lower singularity.

We plot this experimental data, fitting it with straight lines of
different slopes at the two extremes of the data sets. There does
appear to be a change in slope around $\nu=0.5$, in qualitative
agreement with theory.  While the theory does not predict a
discontinuous slope, it does predict a nonanalyticity which resembles
a change of slope. Since the data points are so sparse, we have not
tried a detailed fitting to the predicted functional form. The
magnitude of this nonanalyticity is given by the prefactor of the
$\epsilon^{3/2}$ term in (\ref{discont}). (We note that the
aforementioned background slope from the planes is present in the
experimental data but not in the theoretical curve, so the relative
magnitude of the discontinuity is different.) Experiments using a much
finer sampling of magnetic field values would be required to verify
this prediction.

In this paper we have considered a magnetic field parallel to the
$c$-axis. In the case where it is parallel to the $a-b$ planes there
may be additional interesting anisotropy effects \cite{vekhter}.
Unfortunately, the energy scales seem to be such as to put this out of
the experimentally accessible magnetic field range \cite{junod2}. An
interesting possibility is to consider a tilted magnetic field such
that the energy scales are experimentally accessible while the
in-plane component is still strong enough to yield observable
anisotropy effects. A gap function of mixed symmetry would have a
clear signature on this anisotropy. Another interesting extension is
to consider the role of the paramagnetic response of the electrons to
the magnetic field \cite{para} --- an effect which has been completely
neglected in the present work. While the energy scales seem to be such
as to make this a reasonable assumption, it could well be that for the
in-plane magnetic field, the paramagnetism is of comparable
importance.

We thank I. Vekhter for useful discussions. Research supported in part
by the Natural Sciences and Engineering Research Council (NSERC) and
by the Canadian Institute for Advanced Research (CIAR).


\begin{references} 

\bibitem{dwave} C. C. Tsuei and J. R. Kirtley, Physica C
{\bf 282}, 4 (1997).

\bibitem{stm1} I. Maggio-Aprile et al., Phys. Rev. Lett. {\bf 75},
2754 (1995).

\bibitem{stm2} Ch. Renner et al., Phys. Rev. Lett. {\bf 80}, 3606
(1998).

\bibitem{dv} J. M. Valles et al., Phys. Rev. B {\bf 44}, 11986 (1991).

\bibitem{caroli} C. Caroli, P. G. de Gennes and J. Matricon,
Phys. Lett. {\bf 9}, 307 (1964).

\bibitem{kh} C. K\"ubert and P. J. Hirschfeld, Solid State Comm. {\bf
105}, 459 (1998)

\bibitem{dswave} A. G. Sun et al, Phys. Rev. Lett. {\bf 72}, 2267
(1994); A. S. Katz et al., Appl. Phys. Lett. {\bf 66}, 105 (1995);
A. G. Sun et al., Phys. Rev. B {\bf 54}, 6734 (1996); R. Kleiner et
al., Phys. Rev. Lett. {\bf 76}, 2161 (1996); J. Lesueur et al.,
Phys. Rev. B {\bf 55}, R3398 (1997).

\bibitem{ddom} K. A. Kouznetsov et al., Phys. Rev. Lett. {\bf 79},
3050 (1997).

\bibitem{standard} I. Sch\"urer, E. Schachinger and J. P. Carbotte,
Jour. Low Temp. Phys. {\bf 115}, 251 (1999).

\bibitem{vanhove} C. O'Donovan and J. P. Carbotte, Phys. Rev. B {\bf 55}, 
1200 (1997); ibid {\bf 55}, 8520 (1997).

\bibitem{volovik} G. E. Volovik, JETP Lett. {\bf 58}, 469 (1993).

\bibitem{vekhter} I. Vekhter et al., Phys. Rev. B {\bf 59}, R9023 (1999).

\bibitem{tinkham} see for example, M. Tinkham, {\it Introduction to
Superconductivity, 2nd Ed.}, (McGraw-Hill, New York, 1996).

\bibitem{units} I. Vekhter et al., Proceedings PPHMF-III, cond-mat/9811315. 

\bibitem{moler} K. A. Moler et al., Phys. Rev. Lett. {\bf 73},
2744 (1994); Phys. Rev. B {\bf 55}, 3954 (1997).

\bibitem{fisher} R. A. Fisher et al., Physica C {\bf 252}, 237
(1995). 

\bibitem{junod1} A. Junod et al., Physica C {\bf 282}, 1399 (1997). 

\bibitem{sisson} D. L. Sisson et al., cond-mat/9904131.

\bibitem{maki} K. Maki and M. T. Beal-Monod, Phys. Rev. B {\bf 55}, 
11730 (1997).

\bibitem{junod2} A. Junod, private communication.

\bibitem{para} H. Won, H. Jang and K. Maki, cond-mat/9901252.

\end{references}
\end{document}